\RequirePackage{fix-cm}
\documentclass[twocolumn]{svjour3}          
\smartqed  
\usepackage{graphicx}
\usepackage[outdir=./]{epstopdf}
\usepackage{hyperref}
\usepackage{inputenc}
\begin{document}

\title{On the use of BGP communities for fine-grained inbound traffic engineering
}


\author{Wenqin Shao \and
		Fran\c{c}ois Devienne \and
		Luigi Iannone\and
		Jean-Louis Rougier
}


\institute{
		W. Shao \at
			Telecom ParisTech, Paris, France\\
		  	\email{wenqin.shao@telecom-paristech.fr}
		\and
		F. Devienne \at
			Border 6, Lille, France\\
			\email{francois.devienne@border6.com}
		\and
		L. Iannone \at
			Telecom ParisTech, Paris, France\\
			\email{luigi.iannone@telecom-paristech.fr}
		\and
        JL. Rougier \at
        		Telecom ParisTech, Paris, France\\
			\email{rougier@telecom-paristech.fr}
}

\date{Received: date / Accepted: date}

\maketitle

\begin{abstract}
In the context of Border Gateway Protocol (BGP), inbound inter-domain traffic engineering (TE) remains a difficult problem without panacea. Each of previously investigated method solves a part 
of the problem. 
In this study, we try to complement the map by exploring the use of BGP communities. 
With BGP community based polices enabled in transit provider networks, we are able to manipulate incoming traffic for stub Autonomous System (AS) in a finer granularity than known techniques by customizing the AS-paths perceived by remote networks. 
We analyze the constraints using this technique, along with its effectiveness and granularity. 

\end{abstract}

\section{Introduction}
\label{sec:intro}

The Internet is composed of interconnected Autonomous Systems (AS).
At the time of this writing, there are nearly $49,000$ of them\cite{Cidr}.
Each AS, identified by its AS number (ASN), decides for itself how its own networks are operated.
ASes can be classified into two major types: stub and transit. 
Stub ASes, where most of the Internet traffic is originated and destined, are connected to the rest of the Internet by transit ASes, i.e. transit providers.
The role of Border Gateway Protocol (BGP)\cite{Rfc4271}, the current inter-domain routing protocol, is simply to hold all that much ASes together, thus forming the Internet, regardless of their types.

As AS functions independently, BGP works in a distributive manner:
each AS advertises to its neighbours only the best route, i.e. the route itself uses, for its own prefixes or ones learnt from its neighbours. 
AS can tune its BGP route decision process to enforce local polices. 
For instance,  AS can filter routes either before or after route decision process. 
BGP attributes, such as Local Preference (LP), can also be manipulated so that certain route is privileged over other candidates. 
Consequently, Internet routes are in fact accumulated results of successive individual choices made by ASes on the path from source to destination.
Different from Interior Gateway Protocol (IGP) routes, BGP routes are not necessary shortest in terms of AS-path or router hops.

Inter-domain traffic engineering (TE) is a notion that comes to life with the rise of multi-homed ASes, i.e. ASes, stub or transit, having
several inter-domain links toward one single provider or different providers.
Multi-homed ASes advertises its reachable prefixes via all of its inter-domain links.
A nature question that then occurs is: which transit should be used in sending and receiving traffic so that certain TE objectives, which can be cost or performance related, are achieved.
Techniques involved in egress transit selection are known as outbound TE.
AS has full control when it comes to sending traffic out.
On the other hand, AS doesn't have a grip on incoming traffic which makes inbound TE more difficult by nature.
What can be achieved within in the scope of inbound TE is to influence the route decision process in other ASes,
by manipulating the routes announced.
It has to be noted that there are situations where outbound TE alone is not necessarily sufficient.
For example, more than $50\%$ of inter-domain traffic is of web type\cite{Labovitz2011}, the performance of which depends on Round-Trip Time (RTT) instead of One-Way Delay, which urges that paths on both directions be optimized.
One other case concerns ASes with mainly incoming traffic, e.g. residential Internet Service Providers (ISP).
If they are to balance the traffic load to avoid congestion, manipulating outgoing traffic alone could not be enough.

Realizing the importance of inbound TE, we intend to provide a guide on how to use BGP communities, especially the ones setting transit AS-path prepending, to conduct fine-grained inbound inter-domain TE.
And we focus on the TE needs of stub ASes, as opposite to transit ASes, representing nearly $86\%$ of total ASes in number\cite{Cidr}.
Nonetheless, the methods presented in this paper and the related reasoning can be applied to transit ASes as well.

\section{Related Works}
\label{sec:relate}

In fulfilling following listed TE objectives, an AS should be capable of manipulating incoming and outgoing traffic, i.e. move a certain part of traffic from one link or transit provider to some others, or balance the target traffic over several inter-domain links.
\begin{itemize}
\item Better transmission performance by means of congestion avoidance, load balancing, etc.\cite{Rimondini2013}\cite{Liu2008}\cite{Lee2004}\cite{Goldenberg2004};
\item Minimizing transmission cost\cite{Raja2014}\cite{Goldenberg2004};
\end{itemize}

Outbound TE methods matured over the years, meanwhile inbound TE remains in a less developed situation.
The cause is two-fold. 
First, inbound TE is by nature more difficult than outbound TE.
While an AS can fully decide how it sends out certain traffic, it can not but only influence the route decision of upstream ASes.
Second, inbound TE is sometimes not perceived as necessary,  especially for content providers who mainly care about how to push the content out, since outgoing traffic is more important in volume. 
However, as we argued in $\S$~\ref{sec:intro}, performance of a large part of Internet traffic bases on RTT which takes as well the delay of inbound path into consideration.
Therefore incoming traffic performance is essential to the overall optimization and should not be left behind.

Feamster et al. offered a comprehensive guideline on outbound TE in 2003\cite{Feamster2003}.
Same year, Quoitin et al. succinctly presented several techniques for inbound TE \cite{Quoitin2003}, 
including selective advertisement, AS-path prepending, Multi-Exit Discriminators (MED),
among which BGP communities (RFC 1997\cite{Rfc1997}) appeared to be a promising tool.
Further, Quoitin et al. tried to promote and standardize the use of BGP extended communities (RFC 4360\cite{Rfc4360}) in controlling route propagation known as redistribution communities\cite{Rdcom}\cite{Quoitin2004}.
Mean time, Quoitin et al. investigated majors drawbacks of using BGP communities in controlling route propagation \cite{Quoitin2004a}.
Later on, Donnet et al. reported an increasing use of BGP communities by operators since 2004\cite{Donnet2008}.
All along the road, the main function set of BGP communities remain mainly unchanged: Local Preference (LP) control, BGP route propagation control and transit AS-path prepending.

Basing on these previous works, we try to push the finesse of inbound TE operation to a new level. And the rest of the paper is organized as follows: $\S$~\ref{sec:1} reviews current practices of inbound TE; in $\S$~\ref{sec:2}, we introduce three community based policies in transit networks that make enhanced inbound TE possible for stub ASes; Methods of leveraging these polices to perform fine-grained inbound TE are discussed in $\S$~\ref{sec:3}; $\S$~\ref{sec:concl} concludes this work and outlines future directions.

\section{Current practices of inbound TE}
\label{sec:1}
\begin{figure}[htbp]
\centering
\includegraphics[scale=0.4]{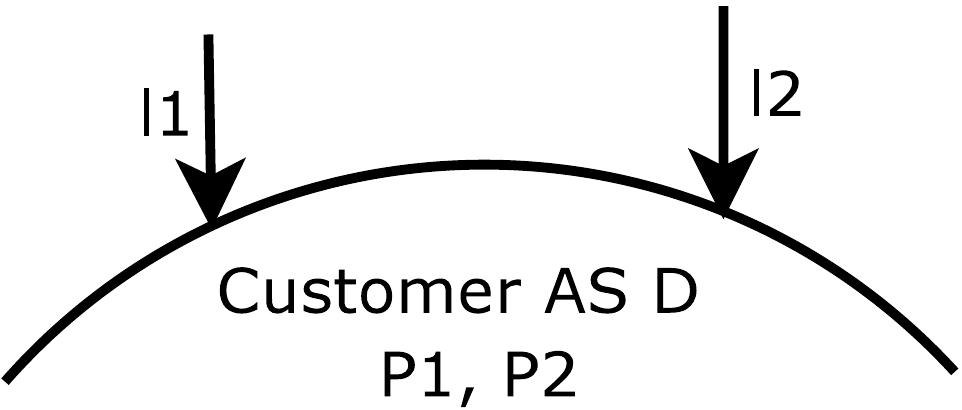}
\caption{A stub AS $D$ \label{fig:0}}
\end{figure}

We give here a brief review on current practices of inbound TE, which mainly come down to the following types:
\begin{itemize}
\item Selective advertisement
\item AS-path prepending
\item MED
\end{itemize}
We demonstrate how these techniques work on stub AS $D$ illustrated in Figure~\ref{fig:0}.
AS $D$ purchases two inter-domain links, $l1$ and $l2$.
Meantime, AS $D$ owns two separate prefixes, \texttt{P1} and \texttt{P2}, and announces routes for them.

\subsection{Selective advertisement}
\label{sec:11}
The essence of this technique is to announce different local prefixes on different links or to different peers.
It can be used to balance incoming traffic or avoid bad performance transit providers.
Suppose that AS $D$ in Figure~\ref{fig:0} wants to balance the traffic coming in over $l1$ and $l2$,
it can announce only \texttt{P1} on $l1$ and \texttt{P2} only on $l2$.
This will force all the traffic whose destination belongs to \texttt{P1} pass via $l1$
and that heading to \texttt{P2} via $l2$.
The major drawback of this solution is that if one of these two links fails, 
relating traffic won't be able to arrive at AS $D$ through the other link.

A variation of this technique is to announce more specific prefixes on corresponding links.
Let's now assume that \texttt{P1'} is a sub-prefix of \texttt{P1} hosting several important servers
for whom a dedicated link is purchased (e.g. $l1$ is used). 
AS $D$ advertises on $l1$ prefix \texttt{P1'} and on $l2$ both \texttt{P1} and \texttt{P2}.
According to longest prefix matching principal, traffic whose destination belongs to \texttt{P1'} reaches AS $D$ preferentially via $l1$.
The advantage of this solution is that \texttt{P1'} is still reachable via $l2$ if $l1$ fails.
The side effect is that it gives rise to BGP route table inflation.

\subsection{AS-path prepending}
\label{sec:12}

With this technique, routes for local prefixes with different AS-path length are advertised on different links or to different peers, allowing a stub AS to indicate its preferences toward its providers. 
It can be used to balance incoming traffic. As argued in \cite{Quoitin2004a}, its granularity is coarse and  effectiveness poor, thus result unpredictable.

Still considering AS $D$ in Figure~\ref{fig:0}, assume that its primary link is $l1$ while $l2$ serves as a live backup.
AS $D$ could announce all his prefixes normally on its primary link, i.e. $l1$,
and prepend its own ASN several times in the AS-path attribute when announcing routes on $l2$.
To be noted that not all the incoming traffic necessarily reaches AS $D$ via its primary link even if prepending is done.
Because, AS-path prepending only modifies the AS-path length seen by upstream ASes, the priority of which is inferior to LP attribute in BGP route decision process.
In order to obtain finer control, this technique can be combined with selective advertisement, i.e. announcing routes for smaller prefixes.

\subsection{MED attribute}
\label{sec:13}

MED is an optional attribute only for AS multi-connected to another AS
to communicate the ranking/preference of links. Link with the smallest MED value has top priority and thus is used for incoming traffic.

In order to achieve the same TE objective described in the previous section, we assume that the same transit provider is on the other end of the two links.
When advertising both prefixes on $l1$ and $l2$, different MED values are used.
The route for \texttt{P1} on $l1$ is attached with a smaller MED value (more preferable) than that on $l2$; the route for $P2$ on $l2$ has smaller MED value than that on $l1$.

It has to be mentioned that the use of MED attribute is normally subject to negotiation between peering ASes. 
Some providers just simply ignore this attribute in its BGP route  decision process.
However, according to \cite{Suv2014}, the use of MED is by no means rare.

\section{Enhanced inbound TE via community attribute}
\label{sec:2}

The limitations of the above mentioned inbound TE techniques are well known.
For selective advertisement, it either suffers from connectivity issues or gives rise to BGP table inflation. 
As for AS-path prepending, it is notorious for unpredictable results.
Furthermore, for both methods, the granularity, defined by the prefix size announced by the local AS, is poor.

Therefore techniques allowing finer granularity and better control are necessary. 
We argue that these objectives can be achieved via ingress community policies supported by transit providers.
Such ingress community policies define the actions taken by transit provider 
on receiving customer routes with certain BGP community attributes attached.
Donnet et al. conducted a survey study on this matter in 2008\cite{Donnet2008}.

BGP community attributes used for this purpose have only local meaning to transit provider, thus non-transitive.
They are removed when routes are announced outside the transit provider network.
Several BGP communities can be attached to a single route,
as long as the the total length of a BGP message does not exceed the limit of $4096$ bytes\cite{Rfc4271}.
A BGP community defined in RFC 1997\cite{Rfc1997} has $4$ bytes for its value field, while extended BGP community defined in RFC 4360\cite{Rfc4360} has 8 bytes.

The actions taken by providers on receiving routes containing communities attributes can be classified into three major kinds: 
\begin{itemize}
\item LP control
\item Route propagation control
\item Transit AS-path prepending
\end{itemize}
We describe in detail these three types of policy.
During the discussion, source AS means always the AS sending traffic, and destination AS is the one where traffic heads and corresponding routes are originated.
We intend to perform inbound TE for destination AS.

\subsection{LP control} 

When a transit provider receives routes from its customer (destination AS), it marks the customer routes with a default LP, which is normally higher than those given to routes received from peers and providers\cite{Gao2001}.
With BGP communities, the customer can set the LP inside the transit provider network.

If the LP assigned to a destination AS originated route is of minor priority to other routes for the same prefix, transit provider will use one of the other routes to forward corresponding traffic.
As a result, traffic no longer comes in via the transit provider with community policy enabled.

Destination AS can use this policy to achieve goals like load balancing or congestion avoidance.
This method can be regarded as an enhanced version of traditional AS-path prepending with more certainty, as LP comes before AS-path in transit provider's route decision process.

\subsection{Route propagation control} 

Customer routes with certain communities won't be announced to
one or a set of peering ASes (transit provider's other customer ASes are normally excluded),
a certain geographical region (EU, for example) or to a certain Internet Exchange Point (IXP).
The granularity of this type of policies varies from provider to provider.
Since this policy won't help much in moving a part of the traffic from one ingress transit provider to some other, we will no further discuss this method.
However it could be extremely effective in filtering unwanted incoming traffic in situations like DDoS attack, as corresponding prefix is simply black-holed to some parts of the Internet.

\subsection{Transit AS-path preprending}

AS transit provider capable of this type of policies can prepend itself several times in the AS-path when advertising received destination (customer) AS routes to certain peers. 
Prepending time and upstream peers to which prepended routes are announced are defined by community values.
Given the fact that several communities can be assigned to a single route,
it is possible to advertise customer routes with different prepending times to different upstream peers,
which in fact expresses destination AS' preference to transit provider's upstream ASes.

What's unique and promising of this method is that it is possible to have the Internet route traffic not only based on traffic destination address but also on traffic source address and thus achieves much more granular TE objectives compared to other methods.

Many big international transit providers are now flattening the AS-level Internet architecture by directly peering with more and more stub ASes\cite{Labovitz2011}, which endows this policy increasing TE value.

\section{Fine-grained inbound TE}
\label{sec:3}

This section demonstrates the general traffic manipulation methods based on transit provider's ingress community policies.

\subsection{Classification by granularity}
\label{sec:31}

TE objectives such as load balancing and congestion avoidance can be realized by moving a part of the traffic from one link to another.
The smallest unit of traffic that we can manipulate without impacting other traffic defines the granularity of the operation.
A BGP flow or a BGP routing object can be identified by the following 4-tuple:
\[
\{ Prefix_{src}, ASN_{src}, Prefix_{dst}, ASN_{dst} \}
\]
The reason why we use prefix along with ASN in identifying TE target traffic is two-fold:
\begin{itemize}
\item BGP routing is prefix based instead of ASN based.
\item Some big AS could have several prefixes and have them distributed in sites that are geographically apart form each other where heterogeneous providers are used.
\end{itemize}

In the following discussions, we use $D$ to denote destination network and its ASN, meaning the destination of the traffic, and accordingly $S$ for source network. AS $D$ advertises routes for its prefixes, toward which AS $S$ generates traffic. We try to perform inbound TE for AS $D$.
All possible TE target traffic can be classified into following types according to their granularity: ($\mathbf{*}$ is a wild card symbol)
\begin{description}
\item[\textbf{Destination prefix based}]\hfill \\ Traffic of this type can be formulated as $\{*,*,*,D\}$ or $\{*,*,P,D\}$.
$\{*,*,P,D\}$ stands for all the traffic destined to a certain prefix \texttt{P} of AS $D$ and is what normal BGP, without community TE, announces routes for. It thus represents the best granularity of current practices.
$\{*,*,*,D\}$ can be seen as the universe of $\{*,*,P,D\}$, standing for all the traffic to the destination AS $D$.
Since there is no interest in moving the entire traffic, this case is no further discussed in destination based TE.
\item[\textbf{Source ASN based}]\hfill \\ Traffic of this type can be formulated as $\{*,S,*,D\}$ or $\{*,S,P,D\}$.
With source ASN based TE, traffic coming from different source ASes may be required to be routed differently. 
As BGP route announcement is prefix based, we discuss $\{*,S,P,D\}$ type of traffic by nature. 
Manipulation of $\{*,S,*,D\}$ type is in fact a composition of $\{*,S,P,D\}$ operations, i.e. applying same configuration to all AS $D$ prefixes for a given source AS $S$.
Source ASN based TE is of better granularity than destination based one, as it differentiates the inbound path not only by destination prefix but also by source AS.
\item[\textbf{Source prefix based}]\hfill \\ Traffic of this type can be formulated as $\{P',S,*,D\}$ or $\{P',S,P,D\}$.
In this case, destination AS wants to distinguish the routing of incoming traffic by its source addresses, in unite of prefix. Due to the fact that an AS can have one or several prefixes, source prefix based TE is of equal or finer granularity than that of source ASN based.
\end{description}

\subsection{Per type Scenarios}
\label{sec:32}

In subsequent sections, we show how to manipulate the above-listed TE objects using community based policies presented in $\S$~\ref{sec:2} and what are the prerequisites, impacts and granularity of the corresponding operations.

\subsubsection{Destination prefix based}
\label{sec:321}

Manipulating this type of traffic can serve for load balancing and congestion avoidance purposes.
Selective advertisement and traditional AS-path prepending discussed in $\S$~\ref{sec:11} and $\S$~\ref{sec:12} achieve this granularity. We demonstrate that LP control policy is of the same granularity.

\begin{figure}[htbp]
\centering
\includegraphics[scale=0.4]{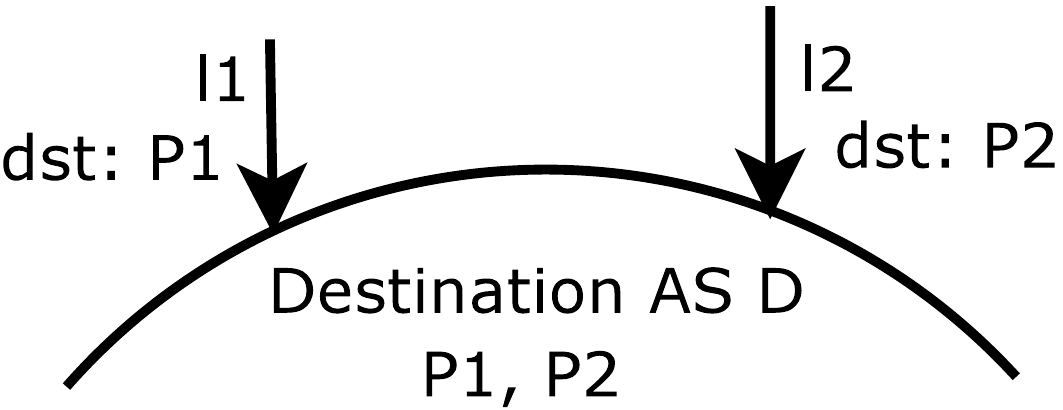}
\caption{Destination prefix based inbound TE \label{fig:1}}
\end{figure}

Let's imagine a customer AS $D$, illustrated in Figure~\ref{fig:1}.
It has two prefixes that do not intersect: \texttt{P1} and \texttt{P2}.
Two separate physical links are purchased for BGP peering: $l1$ and $l2$.
AS $D$ wants that all the traffic to \texttt{P1} enter AS $D$ by $l1$, while that to \texttt{P2} by $l2$, which can be formulated as:
\begin{itemize}
\item On $l1$:
\begin{itemize}
\item $\{*,*,P1,D\}$
\end{itemize}
\item On $l2$:
\begin{itemize}
\item $\{*,*,P2,D\}$
\end{itemize}
\end{itemize}

\paragraph{}
There are two possibilities:
\begin{enumerate}
\item The other ends of the $l1$ and $l2$ belong to a same provider.
\item Two different providers on the other side of $l1$ and $l2$.
\end{enumerate}

\paragraph{Case 1, one same provider:}
As seen in $\S$~\ref{sec:11} and $\S$~\ref{sec:12}, selective advertisement and MED can be employed.
Otherwise, destination AS $D$ can come to the resort to LP control policy.
AS $D$ announces on $l1$ separate routes for \texttt{P1} and \texttt{P2}. 
The route for \texttt{P1} is attached with a community value leading to
a higher LP inside the transit provider that attached to the route for \texttt{P2}.
For example, $100:50$ leads to a LP equals to $50$ inside transit provider, while $100:100$ to a LP equaling to $100$.
Thus $100:100$ is attached to the route for \texttt{P1} and $100:50$ for \texttt{P2} on $l1$.
On $l2$ should be just the reverse : $100:50$ to \texttt{P1} and $100:100$ to \texttt{P2}. These AS $D$ announced routes can be illustrated as follows:
\begin{itemize}
\item On $l1$:
\begin{itemize}
\item Prefix \texttt{P1}: 100:100
\item Prefix \texttt{P2}: 100:50
\end{itemize}
\item On $l2$:
\begin{itemize}
\item Prefix \texttt{P1}: 100:50
\item Prefix \texttt{P2}: 100:100
\end{itemize}
\end{itemize}

\paragraph{Case 2, two different providers:}
In this case, community policies that change the LP within a provider are no longer effective. 
However selective advertisement presented in $\S$~\ref{sec:11} still works.
AS $D$ advertises only the route for \texttt{P1} on $l1$ and only the route for \texttt{P2} on $l2$. 
The side effect is no automatic traffic shift on link failure.

We can also achieve this goal by employing transit AS-path prepending community policy.
\begin{figure}[htbp]
\centering
\includegraphics[scale=0.4]{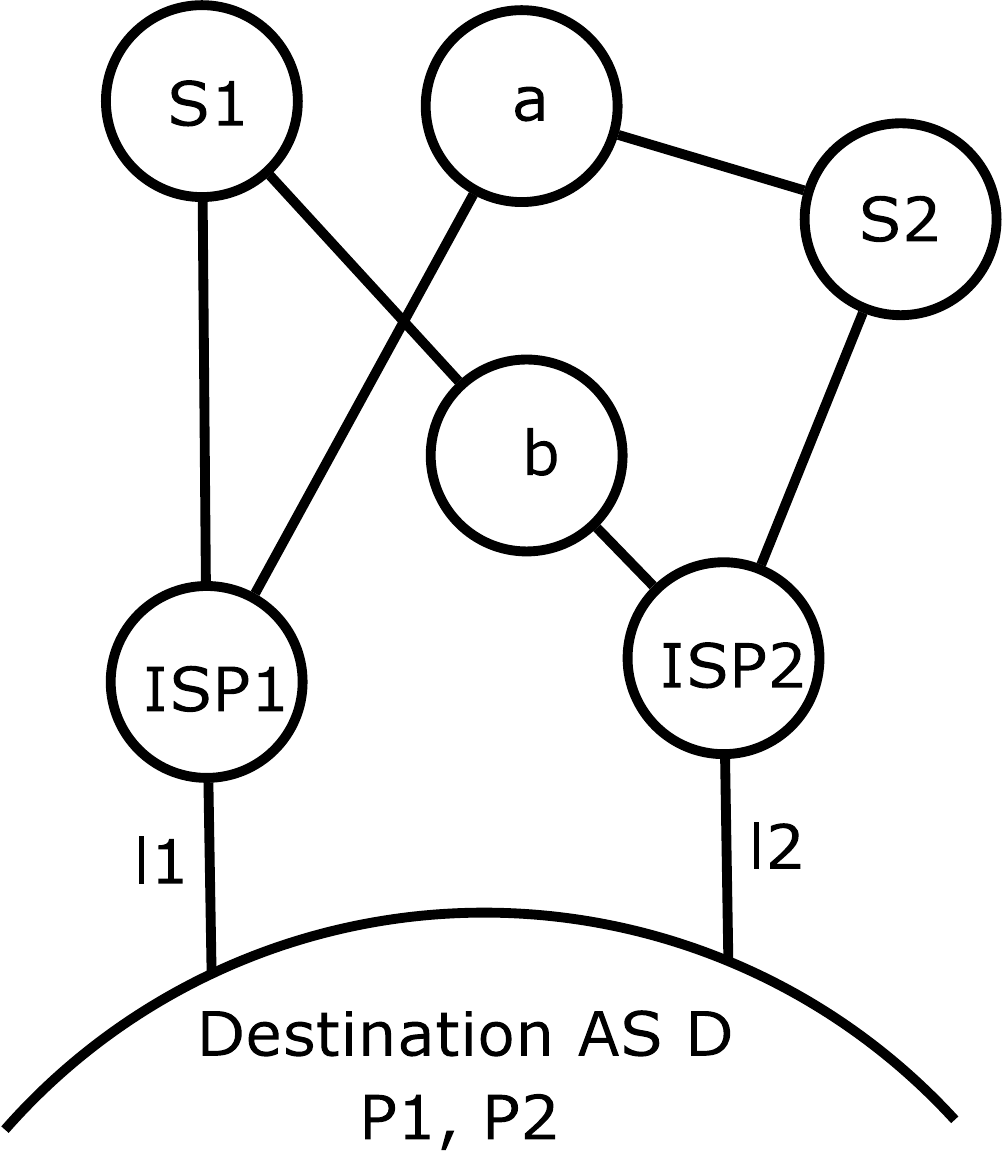}
\caption{client AS $D$ with two different providers ($\#1$) \label{fig:2}}
\end{figure}
Let's consider the topology depicted in Figure~\ref{fig:2} as representative.
In the graph, $ISP1$ and $ISP2$ are two different providers from whom destination AS $D$ purchases its connections.
$S1$ and $S2$ are two possible source ASes.
$S1$ is multi-homed at $ISP1$ and provider $b$, while $S2$ at $ISP2$ and provider $a$.

Without TE,
$S1$ will learn following routes for \texttt{P1} and \texttt{P2}:
\begin{itemize}
\item Prefix: \texttt{P1}
\begin{itemize}
\item AS-path: \texttt{ISP1,D}
\item AS-path: \texttt{b, ISP2, D}
\end{itemize}
\item Prefix: \texttt{P2}
\begin{itemize}
\item AS-path: \texttt{ISP1,D}
\item AS-path: \texttt{b, ISP2, D}
\end{itemize}
\end{itemize}
The one with shorter AS-path will be selected and used, if LP attributes are the same for $ISP2$ and $b$ learnt routes in $S1$. Thus traffic from $S1$ toward both \texttt{P1} and \texttt{P2} of $D$ passes via $l1$.

In order that traffic destined to \texttt{P1} flows through $ISP1$ thus $l1$
and that to \texttt{P2} through $ISP2$ and $l2$, the following should be done:
\begin{itemize}
\item On $l1$, AS $D$ announces route for \texttt{P1} normally
while route for \texttt{P2} attached with a community value
that makes $ISP1$ prepend itself twice when propagating this route to $S1$.
\item On $l2$, routes for \texttt{P1} and \texttt{P2} are announced without special treatment.
\end{itemize}

Consequently, $S1$ will receive two routes with following AS-paths for \texttt{P1}:
\begin{itemize}
\item Prefix: \texttt{P1}
\begin{itemize}
\item AS-path: \texttt{ISP1, D}
\item AS-path: \texttt{b, ISP2,D}
\end{itemize}
\end{itemize}
The route traversing $ISP1$ will be chosen supposing that the two routes have same LP inside $S1$.

For \texttt{P2}, $S1$ will have the following options:
\begin{itemize}
\item Prefix: \texttt{P2}
\begin{itemize}
\item AS-path: \texttt{ISP1, ISP1, ISP1, D}
\item AS-path: \texttt{b, ISP2, D}
\end{itemize}
\end{itemize}
The route traverses $ISP2$ will be the chosen one.

Let's now consider the $S2$ as a possible source AS.
Obviously the prepending pattern will be different to achieve the TE goal:
prepending $ISP2$ twice when propagating the route for \texttt{P1} to $S2$.

This example demonstrates that in order to manipulate destination prefixes based traffic via AS-path prepending communities,
each source AS may require a different prepending pattern. 
Besides this, it has other two important constraints:
\begin{enumerate}
\item LP must be the same for routes in all upstream ASes so that AS-path attribute could be attended in route decision process. Destination networks $D$, for whom inbound TE is performed, has no absolute control on this matter.
In practice, AS $S$ may use their own outgoing policy altering LP attribute. Thus AS-path length is not considered.
\item The operation requires visibility on BGP route information database of source AS,
or sufficient, up-to-date knowledge on the AS-level Internet topology, both of which are quite unavailable, to decide the prepending pattern.
\end{enumerate}
Given these two constraints, a closed-loop mechanism to tryout all possible prepending patterns would be necessary.
However this may lead to a long route convergence time and route flapping.

Selective advertisement, MED (constrained to same transit provider), LP control community and transit AS-path prepending community are all capable of manipulating destination prefix based traffic. 
MED and LP control community can impact target traffic with certainty. 
If combined with selective advertisement, i.e. splitting \texttt{P} into smaller pieces, finer granularity can be achieved. 
However they still can not move traffic that originated from a specific source network without touching the rest. 
AS-path prepending community, though bounded by many constraints, shows potential for even finer manipulation, with source AS taken into consideration, as we will show in the following section.

\subsubsection{Source ASN based}
\label{sec:322}

\begin{figure}[htbp]
\centering
\includegraphics[scale=.4]{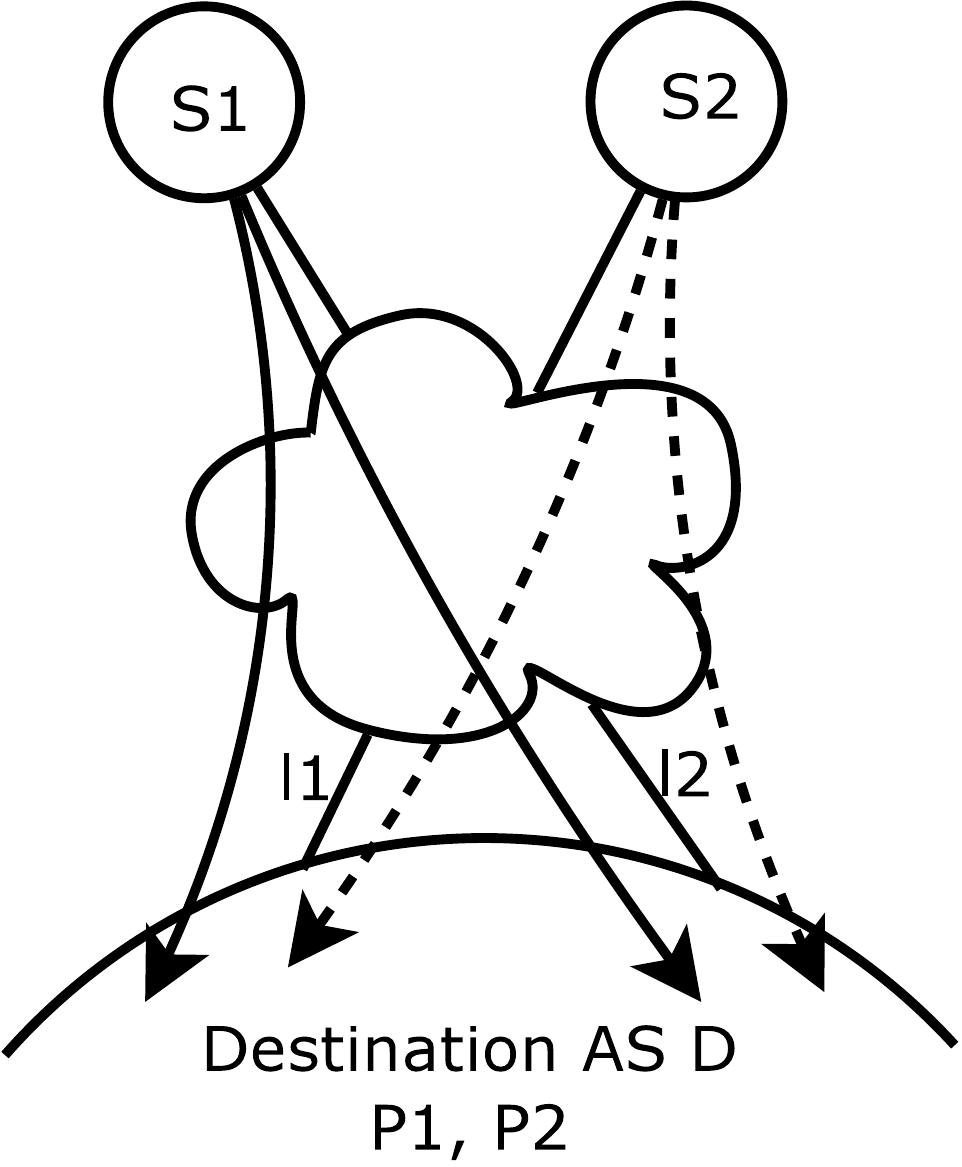}
\caption{Source ASN based Inbound TE\label{fig:3}}
\end{figure}

In moving this kind of traffic, we can imagine a scenario like this:

Customer AS $D$ has two physical links, $l1$ and $l2$, for BGP peering, as illustrated in Figure~\ref{fig:3}.
Two separate prefixes, \texttt{P1} and \texttt{P2}, belong to AS $D$.
There are two possible source ASes, $S1$ and $S2$.
The target traffic pattern desired by AS $D$ is as follows:
\begin{itemize}
\item On $l1$:
\begin{itemize}
\item $\{*, S1, P1, D\}$
\item $\{*, S2, P2, D\}$
\end{itemize}
\item On $l2$:
\begin{itemize}
\item $\{*, S1, P2, D\}$
\item $\{*, S2, P1, D\}$
\end{itemize}
\end{itemize}
Traffic from $S1$/$S2$ is balanced on both links, according to its destination prefix.
Furthermore, traffic to \texttt{P1}/\texttt{P2} is as well distributed on both links, depending on its source AS.

\paragraph{}
Again we have two possibilities, just the very same ones as the previous scenario:
\begin{enumerate}
\item The other ends of the $l1$ and $l2$ belong to a same provider.
\item Two different providers at the ends of $l1$ and $l2$.
\end{enumerate}
\paragraph{Case 1, one same provider:}
In this situation, the solution would be out of the scope of BGP if not utterly impossible.
This is because that BGP route decision process bases on destination prefix,
and AS $D$'s transit provider can not distinguish these two kinds of traffic: $\{*,S1,P,D\}$ and $\{*,S2,P,D\}$.
We can further generalize this conclusion: if there is an upstream AS in common in target AS-paths, prepending policy is no longer effective.

\paragraph{Case 2, two different providers:}
Having two different providers is not a guarantee that prepending will work as we wish.
The generalized upstream-AS-in-common constraint still applies.
For instance, two tier 2 providers peered with the destination AS share a very same tier 1 upstream AS and target AS-paths for different source ASes all pass by the tier 1 provider.
In that case, no matter what prepending pattern applied in these two tier 2 ISPs,
only one of them will be chosen by the tier 1 in forwarding traffic to prefixes of destination AS.
Therefore, multi-homing via heterogeneous providers whose up-stream ASes have less intersection has more TE values than that via similar ones.
Apart from this, prepending is capable of moving incoming traffic and we retake the topology depicted in Figure~\ref{fig:2}, and show how to use prepending community to realized the above expressed traffic pattern.

As discussed in $\S$~\ref{sec:321}, the traffic that flows over $l1$ and $l2$ is as follows, when no TE has been performed on Figure~\ref{fig:2}:
\begin{itemize}
\item On $l1$:
\begin{itemize}
\item $\{*,S1,P1,D\}$
\item $\{*,S1,P2,D\}$
\end{itemize}
\item On $l2$:
\begin{itemize}
\item $\{*,S2,P1,D\}$
\item $\{*,S2,P2,D\}$
\end{itemize}
\end{itemize}
All traffic from $S1$ concentrates on $l1$, while that from $S2$ on $l2$. In order to achieve the target traffic pattern, we should move $\{*,S1,P2,D\}$ from $l1$ to $l2$ and $\{*,S2,P2,D\}$ from $l2$ to $l1$. One possible solution for this traffic manipulation can be:
\begin{itemize}
\item Prepend $ISP1$ twice when advertising \texttt{P2} to $S1$.
\item Prepend $ISP2$ twice when advertising \texttt{P2} to $S2$.
\end{itemize}
With this prepending pattern realized by attaching corresponding BGP communities to the routes announced by AS $D$, $S1$ will receive following routes:
\begin{itemize}
\item Prefix \texttt{P1}:
\begin{itemize}
\item AS-path: \texttt{ISP1, D}
\item AS-path: \texttt{b, ISP2,D}
\end{itemize}
\item Prefix \texttt{P2}:
\begin{itemize}
\item AS-path: \texttt{ISP1, ISP1, ISP1, D}
\item AS-path: \texttt{b, ISP2,D}
\end{itemize}
\end{itemize}
$S1$ will therefore chose provider $b$, consequently $l2$, in forwarding traffic toward \texttt{P2}, if the routes learnt from $ISP1$ and provider $b$ are of same LP inside $S1$.

When it comes to $S2$, following routes will be received:
\begin{itemize}
\item Prefix \texttt{P1}:
\begin{itemize}
\item AS-path: \texttt{ISP2, D}
\item AS-path: \texttt{a, ISP1,D}
\end{itemize}
\item Prefix \texttt{P2}:
\begin{itemize}
\item AS-path: \texttt{ISP2, ISP2, ISP2, D}
\item AS-path: \texttt{a, ISP1,D}
\end{itemize}
\end{itemize}
Similarly, $S2$ now chooses provider $a$ and thus $l1$ to send traffic to \texttt{P2}.

\begin{figure}[htbp]
\centering
\includegraphics[scale=.4]{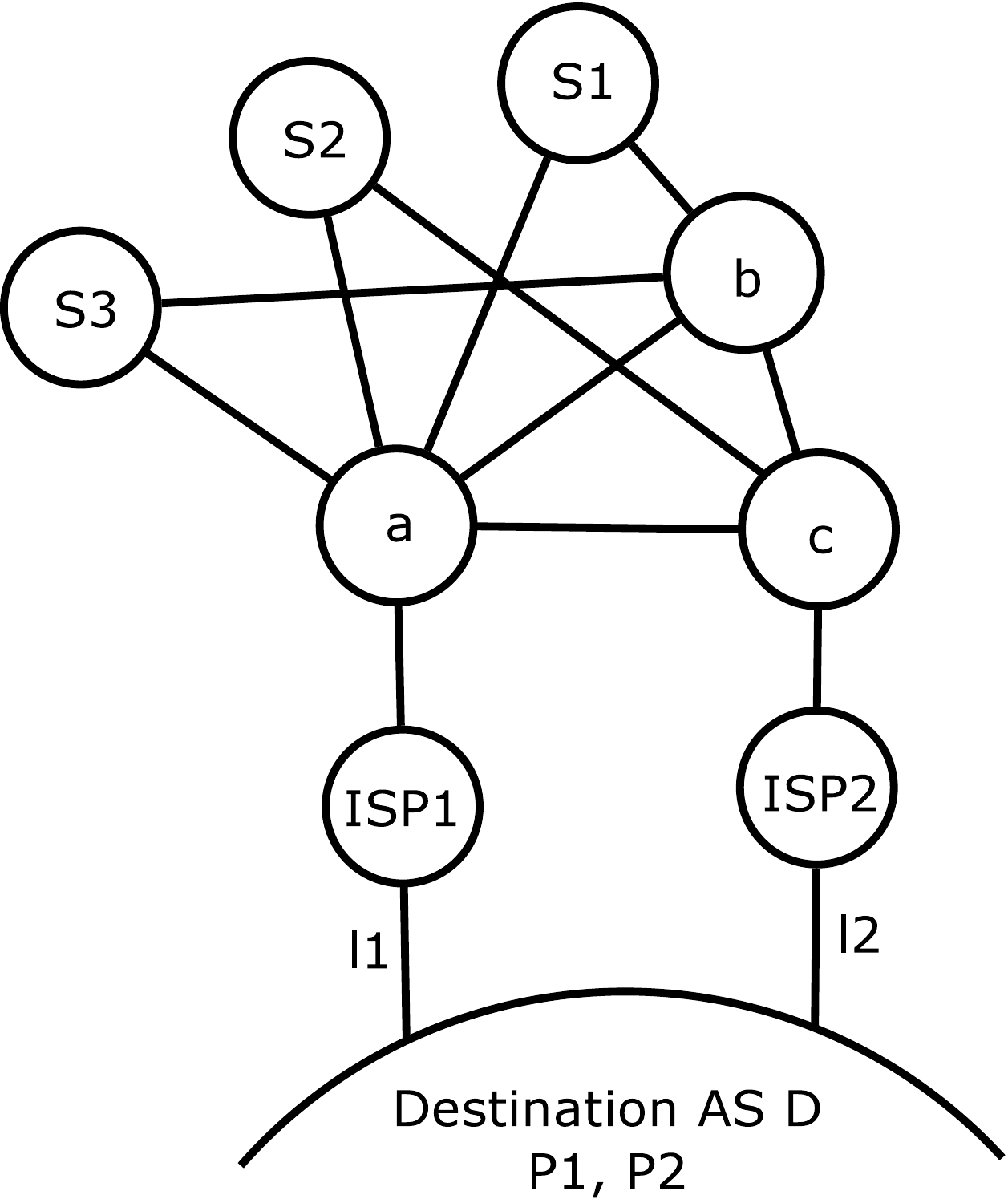}
\caption{client AS $D$ with two different providers ($\#2$) \label{fig:5}}
\end{figure}

The above simple scenario does help to explain the general ideal of inbound TE using prepending community policy and how we arrived at manipulating source ASN based traffic. 
However it fails to reveal the limitations of this operation. 
Let's consider the topology shown in Figure~\ref{fig:5}, which is a little bit more complicate.

In this topology, destination AS $D$ has two BGP connections from $ISP1$ and $ISP2$ and two separate prefixes, \texttt{P1} and \texttt{P2}.
We consider three possible source ASes, $S1$, $S2$ and $S3$.
AS $D$ requires that the traffic from $S1$ enters in following manner:
\begin{itemize}
\item On $l1$:
\begin{itemize}
\item $\{*,S1,P1,D\}$
\end{itemize}
\item On $l2$:
\begin{itemize}
\item $\{*,S1,P2,D\}$
\end{itemize}
\end{itemize}

Without TE, $S1$ receives the following routes to prefixes originated from AS $D$:
\begin{itemize}
\item Prefix \texttt{P1}:
\begin{itemize}
\item AS-path: \texttt{a, ISP1, D}
\item AS-path: \texttt{b, c, ISP2, D}
\end{itemize}
\item Prefix \texttt{P2}:
\begin{itemize}
\item AS-path: \texttt{a, ISP1, D}
\item AS-path: \texttt{b, c, ISP2, D}
\end{itemize}
\end{itemize}
As a result, $S1$ chooses provider $a$ as egress transit for  all traffic destined to AS $D$ under the hypothesis that routes learnt from provider $a$ and $b$ are of same LP inside $S1$. In order to achieve the desired traffic pattern, we can make $ISP1$ prepend itself twice when advertising \texttt{P2} to its upstream AS $a$. 
$S1$ now receives:
\begin{itemize}
\item Prefix \texttt{P1}:
\begin{itemize}
\item AS-path: \texttt{a, ISP1, D}
\item AS-path: \texttt{b, c, ISP2, D}
\end{itemize}
\item Prefix \texttt{P2}:
\begin{itemize}
\item AS-path: \texttt{a, ISP1, IPS1, ISP1, D}
\item AS-path: \texttt{b, c, ISP2, D}
\end{itemize}
\end{itemize}
Consequently, $l2$ becomes the link responsible for incoming traffic toward \texttt{P2}.

However, the odds are high that traffic identified as $\{*,S2,P2,D\}$ and $\{*,S3,P2,D\}$ arrives at AS $D$ via $l2$ as well after prepending, which could be unwanted, since these two source ASes share the same transit provider $a$ from which they receive prepended routes for \texttt{P2} advertised by $ISP1$. 

This example shows clearly the side effects and the granularity of this operation,  which depend a lot on the position of upstream ASes to which prepended routes are advertised.
Nonetheless source ASN based TE is of finer granularity than that of destination prefix based, as certain level of selection is posed on transit providers' upstream ASes.
Again, the manipulation bases on the prerequisite that AS-level topology is known to destination AS and source AS won't distinguish routes from different transit providers by LP attribute.

\subsubsection{Source prefix based}
\label{sec:323}

Source prefix based TE, for example making $\{P1,S,P3,D\}$ and $\{P2,S,P3,D\}$ following two different AS-paths, implies that source network $S$ knows how to route traffic differently according to its source IP and that this difference can be expressed in BGP route decision process. As we know that traditional IP routing and forwarding is destination address based, manipulating type source prefix based traffic is thus not a common practice under current networking framework. However, IETF working group SPRING\cite{SPRING}, acronym for Source Packet Routing in Networking, proposes standards allowing, but not limited to, source address based routing, which could be a clue for source prefix based inbound TE.

\section{Conclusion}
\label{sec:concl}
In this work, we first reviewed some well known inbound TE methods: selective advertisement, AS-path prepending and MED.
The latter two techniques can be combined with selective advertisement to achieve finer granularity in manipulation, but still rest effective only for destination prefix based traffic. Plus AS-path prepending is known for not being an effective measure, while MED is only useful in front of a very same transit provider.
 
In search of alternatives for inbound TE, we investigated
three BGP community-based TE policies adopted by transit providers, route propagation control, LP control and transit AS-path prepending.

Ingress community policies that control route propagation can effectively black-hole certain prefixes of the destination network to a part of the Internet, which could be useful for DDoS mitigation.

BGP communities that set LP values for routes toward destination network within transit provider's network are powerful in moving the entire traffic to certain destination prefixes from one provider to some others, thus effective on destination prefix based traffic.
This method can be used for congestion avoidance and load balancing purposes.
It is supposed to work with more certainty than traditional AS-path prepending done by destination network, as LP is of superior priority to AS-path in transit provider's BGP route decision process. 
Poor granularity is however the short-board of this method. 
With or without selective advertisement, it will impact the entire traffic toward to the relating prefix, the route of which is attaches with corresponding BGP communities.

Looking for finer granularity, we focused latter on BGP communities that allow AS-path prepending that is specific to transit provider's upstream ASes. This allows to set up TE polices not only based on destination prefixes, but also on source AS as well, i.e. source ASN based traffic, at the cost of calculating a pre-prending pattern for each possible source AS with the certain constraints.
The effective granularity, finer than that of destination prefix based TE in worst case, depends largely on the upstream ASes of the transit provider to which prepending is applicable.

An automated and closed-loop mechanism would be required for the calculation or enumeration of prepending patterns. 
Possible route flapping and longer convergence delay should also be attended in prepending processing, which calls for further work.

In practice, we have also noticed that the implementation of ingress community policies varies a lot from provider to provider.
Some of them provide prepending options of fine granularity, some others just simply drop BGP UPDATE messages containing community attributes.
We recommend that this difference should be considered in the choice of providers for multi-homed stub ASes.


\bibliographystyle{ieeetr}
\bibliography{bgp}

\end{document}